\documentclass{Interspeech2024}

\usepackage{booktabs} % For professional looking tables
\usepackage{multirow} % For multi-row cells in tables

\interspeechcameraready 

\title{Dataset-Distillation Generative Model for Speech Emotion Recognition}

\name[affiliation={1,2}]{Fabian}{Ritter-Gutierrez}
\name[affiliation={3,4}]{Kuan-Po}{Huang}
\name[affiliation={2}]{Jeremy H.M}{Wong}
\name[affiliation={1,5}]{Dianwen}{Ng}
\name[affiliation={3}]{Hung-yi}{Lee}
\name[affiliation={1,2}]{Nancy F.}{Chen}
\name[affiliation={1}]{Eng-Siong}{Chng}

\address{
  $^1$Nanyang Technological University, Singapore $^2$Institute for Infocomm Research (I2R), Singapore $^3$National Taiwan University, Taiwan $^4$ASUS Intelligent Cloud Services, Taiwan   $^5$Speech Lab of DAMO Academy, Alibaba Group, Singapore}

\email{s220064@e.ntu.edu.sg, stufarg@i2r.a-star.edu.sg}

\keywords{self-supervised learning, dataset distillation, speech emotion recognition, generative adversarial network.}

\begin{document}

\maketitle

\begin{abstract}
Deep learning models for speech rely on large datasets, presenting computational challenges. Yet, performance hinges on training data size. Dataset Distillation (DD) aims to learn a smaller dataset without much performance degradation when training with it.  DD has been investigated in computer vision but not yet in speech. This paper presents the first approach for DD to speech targeting Speech Emotion Recognition on IEMOCAP. We employ Generative Adversarial Networks (GANs) not to mimic real data but to distil key discriminative information of IEMOCAP that is useful for downstream training. The GAN then replaces the original dataset and can sample custom synthetic dataset sizes. It performs comparably when following the original class imbalance but improves performance by 0.3\% absolute UAR with balanced classes. It also reduces dataset storage and accelerates downstream training by 95\% in both cases and reduces speaker information which could help for a privacy application.
\end{abstract}

\section{Introduction}\label{sec:introduction}
%\vspace{-0.1cm}
End-to-end (E2E) machine learning and self-supervised learning (SSL) techniques have revolutionized speech processing in various tasks \cite{hubert,wavlm,SSLreview}. However, they rely on large data resources for training, posing storage and data processing challenges. For example, \cite{owsm} utilized 180k hours of labelled data and required 20 days of training on 64 GPUs to train a single model. Such data-intensive models present financial and logistical challenges when faced with limited resources while posing severe environmental impact \cite{energynlp}. Despite these issues, the current training paradigm necessitates a vast amount of data\cite{scaling_law_asr}.

Dataset Distillation (DD) \cite{DDreview} has emerged, showing great promise for reducing training costs. DD aims to learn discriminative and informative samples and form a smaller synthetic dataset hoping to retain as much performance as the original dataset. DD deviates from the ``data-selection" paradigm \cite{retrieve_coreset} where a smaller dataset is created by selecting representative data points in the dataset. In contrast, DD learns abstract representations that convey the dataset's most discriminative information, which may or may not look realistic. 

DD is a popular emerging paradigm in Computer Vision (CV) \cite{dd_initial,dc_gradient_matching_2020,dd_distribution_matching,dd_cafe,zhou2023dataset} yet it has not been explored for speech processing tasks. DD for speech processing introduces unique challenges due to the inherent differences between speech signals and images. Speech is a temporal signal with temporal dependencies. Hence, there is relevant information to distil across time.  This paper proposes a first attempt of DD on speech processing task, aiming to 1) significantly reduce the disk storage requirement compared to the original dataset, 2) reduce training time computation on the downstream task, 3) make speaker identity harder to recover to enhance privacy and 4) alleviate data-label imbalance. Such goals should be achieved without considerably hurting downstream model performance when training with the distilled dataset.

Speech emotion recognition (SER) task in the IEMOCAP dataset \cite{iemocap} is chosen as a case study for the following reasons. First, SER is an utterance-level classification task, where the variable length speech sequence is mapped into a single vector for classification. This is a favorable starting point to analyze the feasibility of this research direction on speech processing before extending the approach to speech tasks that make predictions over frames of the speech sequence. Second, while utterance-level classification makes the task more manageable, the subtleties needed to model emotions are challenging and interesting. The DD algorithm will need to convey discriminative information of a speech signal for ER classification. 

\begin{figure}[!th]
  \centering
  \includegraphics[width=1\linewidth]{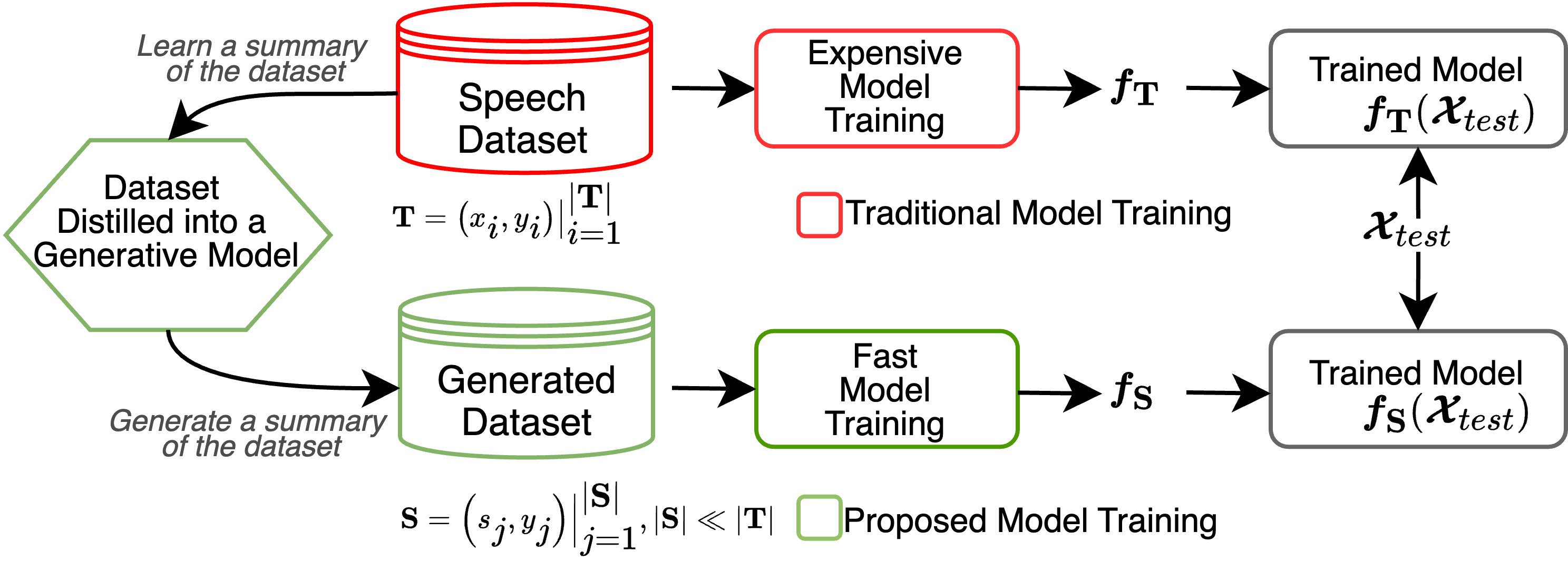}
  \caption{Usage scenario for DD on speech processing tasks. $\boldsymbol{f}_{\mathbf{T}}(\boldsymbol{x}_{test})$ represents inference on a downstream model $\boldsymbol{f}$ trained under dataset $\mathbf{T}$.}
  \label{fig:scenario_usage}
  %\vspace{-0.1cm}
\end{figure}

Fig. \ref{fig:scenario_usage} shows the usage scenario of the proposed method. Rather than training a downstream model with the original dataset, which requires expensive model training due to hyperparameter tuning, downstream architecture selection, and so on, we propose to learn a distribution that summarizes the training data, and that is controlled only by the emotion class labels. By learning a distribution that summarizes the training data across emotion labels, we do not need to retain a record of the original speech. Hence, our proposal implicitly enhances privacy. Nonetheless, this does not means the proposal guarantees private generated representations. Once this summary distribution is learned as a generative model, a custom budget of samples per class can be generated to train downstream models, perform parameter tuning and so on. While training a generator incurs a cost, our proposal aims to provide a generator that replaces the dataset, meaning that training the generator is a once-for-all process.

\begin{figure}[!th]
  \centering
  \includegraphics[width=1\linewidth]{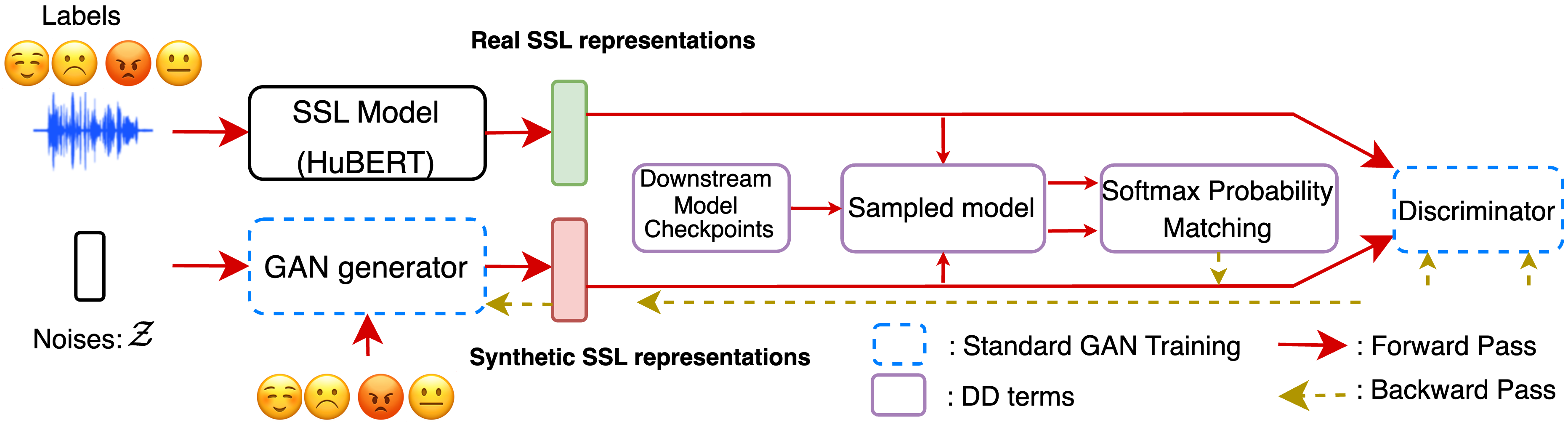}
  \caption{Schematic representation of the proposed DD. The blue dashed lines represent the standard training of a GAN}
  \label{fig:dd_gan_overview}
  \vspace{-0.1cm}
\end{figure}

The method, depicted in Fig \ref{fig:dd_gan_overview}, employs a Generative Adversarial Network (GAN) for DD in IEMOCAP, favored over a Diffusion Probabilistic Model (DPM) due to its smaller size, higher computational efficiency, and quicker on-the-fly data generation capabilities \cite{voice_conversion_diffusion}. Nonetheless, GANs have been designed to generate real-looking data, differing from our goal of learning a summary distribution of the dataset useful for downstream training. Hence, to make the GAN learn discriminative information useful for downstream performance, we propose to bias the GAN by adding a term that minimizes the Kullback-Leibler (KL) divergence between the softmax probabilities of emotion classes of downstream forward passes between the real and synthetic data. We prevent the GAN from merely memorizing the softmax probability distribution by sampling from a variety of downstream model checkpoints, thereby introducing a range of possible KL divergence targets. Furthermore, a diversity penalty term is added to make the GAN sample more diverse data on smaller synthetic dataset sizes. To test the efficiency of the proposed method, we do ablations to see real data test set performance on IEMOCAP. The results obtained show that our proposal consistently maintains close accuracy performance comparable to a model trained on the real IEMOCAP dataset and it is consistently better than a GAN \cite{wassgangp} trained without our proposed criteria with statistical significance at a p-value of 0.05. The proposed method reduces the dataset size and training time by 95\% with minimal performance degradation. Additionally, it improves SER over the real data training when our method samples balanced datasets. Hence, the proposal alleviates data imbalance issues inherent in IEMOCAP. Finally, this proposal implicitly decreases speaker identity information which fosters possibilities for privacy related applications.

\vspace{-0.1cm}
\section{Related Work}

While there is no work directly aiming to distill a dataset for SER or any speech processing task, there is work that leverages GANs for data augmentation in SER. In \cite{gan_feature_dim_2018}, an unconditional and conditional GAN was trained for the IEMOCAP dataset. \cite{data_aug_gan_2019}, uses a conditional GAN to do mel-spectrogram augmentation to improve performance on less representative emotion classes for IEMOCAP.\cite{cyclegan_transfer_emotion_2019} investigates CycleGAN for emotion style transfer, aiming to generate realistic emotion data. The study adds an evaluation of real test sets for models trained on synthetic data only, revealing a performance gap above 8\% between training on real versus synthetic data. There are more similar works using GANs for data augmentation such as \cite{Yi2019AdversarialDA,Yi2022ImprovingSE,related_work_improved_stargan_2021} but with different GANs architectures and some recent work has attempted speech emotion recognition data augmentation using denoising diffusion probabilistic models (DDPM) \cite{diffusion_schuller}.

\section{Dataset Distillation}

Generally speaking, DD aims to learn a small dataset that achieves comparable performance to the original dataset that it is distilling from. Let  $\mathcal{\mathbf{T}}=\left.\left(\mathbf{x_i}, \mathbf{y_i}\right)\right|_{i=1} ^{|\mathcal{\mathbf{T}}|}$ be the real dataset consisting on data-label pairs $\mathbf{x}_i, \mathbf{y}_i$ with $\mathbf{x}_i \in \mathbb{R}^d$ with $d$ the feature dimension and $\mathbf{y}_i \in \mathbb{R}^c$ with $c$ the number of classes. DD aims to create a synthetic dataset $\mathcal{\mathbf{S}}=\left.\left(\mathbf{s}_j, \mathbf{y}_j\right)\right|_{j=1} ^{|\mathcal{\mathbf{S}}|}$ (${|\mathcal{\mathbf{S}}|} \ll {|\mathcal{\mathbf{T}}|}$). Once $\mathcal{\mathbf{S}}$ is learned, the dataset is deployed to train a downstream model, and that model is evaluated on the real data test set. 

DD methods are based on three strategies \cite{DDreview}: i) performance matching: monitoring performance achieved by a neural network with the original dataset versus the synthetic dataset \cite{dd_initial}; ii) gradient matching: match the gradient in a neural network of the original and synthetic dataset at each iteration \cite{dc_gradient_matching_2020}; iii) distribution or feature matching: match the features produced on a neural network for the real and synthetic data \cite{dd_cafe,dd_distribution_matching}. In general, the algorithm will define a fixed budget of number of elements per class when doing DD. Hence, if a different budget is needed, a whole DD training must be done again. Differently, the works \cite{zhao2022synthesizing,dim} distill CV datasets into a generative model. Hence, rather than directly learning a dataset $\mathbf{S}$, they learn a generator $\boldsymbol{g}$ that can sample different datasets based on a sample per class budget. Our proposal is motivated by these ideas.

\section{Dataset Distillation for Speech Emotion}

In this paper, rather than directly learning a dataset $\mathcal{\mathbf{S}}$, a generative model $\boldsymbol{g}$ is learnt to generate summary distributions of $\mathcal{\mathbf{T}}$. Once $\boldsymbol{g}$ is learnt, custom-defined samples per emotion can be generated. Small-size generative models are designed, thereby significantly decreasing storage requirements of the original dataset as seen in Table \ref{tab:disk_storage_dd}. 

The proposed approach consists of two stages, the first stage is a standard GAN training, particularly the conditional Wasserstein GAN implementation with gradient penalty (WGAN-GP) \cite{wassgangp}. In WGAN-GP, the discriminator  $\boldsymbol{d}_\omega$, with $\omega$ the weight parameters, is optimized as,
\begin{equation}
 \resizebox{.91\linewidth}{!}{$
            \displaystyle
\text{L}_{\text {D}_\text{ADV}}(w) =  \underset{\boldsymbol{z} \sim P(\boldsymbol{z} )}{\mathbb{E}}  \left[\boldsymbol{d}_\omega\left(\boldsymbol{g}_\phi(\boldsymbol{z})\right)\right]
- \underset{\boldsymbol{x} \sim P(\boldsymbol{x})}{\mathbb{E}} \left[\boldsymbol{d}_\omega(\boldsymbol{x})\right] + \lambda_1 \text{L}_{\text {GP}} (w),
$}
\label{eq:disc_wgangp}
\end{equation}

\noindent where $\boldsymbol{g}_\phi$ is the generator parametrized by the weights $\phi$, $P(\boldsymbol{z})$, $P(\boldsymbol{x})$ denotes the distribution of noise (latent) vectors and real samples respectively. A noise vector $\mathbf{z}\sim P(\boldsymbol{z} )$ contains the information of the label $\mathbf{y}$ in the form of a one-hot vector, i.e. $\mathbf{z} \equiv [\mathbf{y} \oplus \mathbf{e}]$, with $\oplus$ the concatenation operation and $\mathbf{e} \sim \mathcal{N}(0,1)$.

The gradient penalty $\text{L}_{\text {GP}}(w)$ is needed to have a valid Wasserstein distance computation and $\lambda_1$ controls the importance of this term. We use $\lambda_1=10$ as in the original WGAN-GP \cite{wassgangp}.

The generator in WGAN-GP is trained to minimize,
\begin{equation}
\text{L}_{\text{G}_\text{ADV}}(\phi)= -  \underset{\boldsymbol{z} \sim P(\boldsymbol{z})}{\mathbb{E}}\left[\boldsymbol{d}_w\left(\boldsymbol{g}_\phi(\boldsymbol{z})\right)\right].
\label{eq:gen_wgangp}
\end{equation}

\noindent Additionally, motivated by speech processing research on mel-spectrogram inversion \cite{melgan,hifigan}, we add a feature matching (FM) loss, shown to improve stability for generator training. The FM loss is defined as,%\vspace{-0.05cm}
\begin{equation}
 \resizebox{.91\linewidth}{!}{$
            \displaystyle
\text{L}_{\text{FM}}\left(\boldsymbol{g}_\phi, \boldsymbol{d}_w \right)=\underset{ \substack{\boldsymbol{x} \sim P(\boldsymbol{x})\\ \boldsymbol{z} \sim P(\boldsymbol{z})}}{\mathbb{E}}\left[\sum_{l=1}^M \frac{1}{M} \left|  \boldsymbol{d}_w^{(l)}(\boldsymbol{x} )- \boldsymbol{d}_w^{(l)}( \boldsymbol{g}_\phi(\boldsymbol{z})  ) \right|   \right],
$}
\label{eq:feature_matching}
\end{equation}

\noindent where $\boldsymbol{d}_w^{(l)}$ is the feature map at the ``l-th" layer of the discriminator $\boldsymbol{d}_w$, and $M$ the number of layers. Eq. \eqref{eq:feature_matching} helps the generator to sample features on the same space than the real data.

Finally, to use the conditioning class label information, a cross-entropy loss is added. Then, the final loss for the discriminator is,
\begin{align}
\text{L}_{\text{D}} &= \text{L}_{\text{D}_\text{ADV}}(\omega)  + \lambda_2 \underset{\boldsymbol{x} \sim P(\boldsymbol{x})}{\mathbb{E}}\left[\text{CE}(\boldsymbol{d}_w^{\text{ class}}(\boldsymbol{x}), \boldsymbol{y})\right] \notag \\
&+ \lambda_3 \underset{\boldsymbol{z} \sim P(\boldsymbol{z})}{\mathbb{E}}\left[\text{CE}(\boldsymbol{d}_w^{\text{ class}}(\boldsymbol{g}_\phi(\boldsymbol{z})), \boldsymbol{y})\right],
\label{eq:final_disc_before_dd}
\end{align}

\noindent with $\text{CE}$ denoting the cross-entropy loss, $\boldsymbol{d}_w^{\text{ class}}(\cdot)$ the logits distribution of the emotion classes $\boldsymbol{y}$, and $\lambda_i$ represent scalar weights. The final generator loss is,
\begin{equation}
\resizebox{.91\linewidth}{!}{$
\text{L}_{\text{G}} = \text{L}_{\text{G}_\text{ADV}}(\phi) + \lambda_3 \underset{\boldsymbol{z} \sim P(\boldsymbol{z})}{\mathbb{E}}\left[\text{CE}(\boldsymbol{d}_w^{\text{ class}}(\boldsymbol{g}_\phi(\boldsymbol{z})), \boldsymbol{y})\right] + \lambda_4 \text{L}_{\text{FM}}.
\label{eq:final_gen_before_dd}$}
\end{equation}

\noindent Eq. \eqref{eq:final_disc_before_dd} and \eqref{eq:final_gen_before_dd} are designed to generate data that resembles real instances as done in previous work \cite{data_aug_gan_2019,cyclegan_transfer_emotion_2019}. The aim of this paper diverges from conventional uses of GANs for creating real-looking data. Instead, the focus is on harnessing GANs to generate key discriminative information that serves downstream model performance so that it can be used for DD. This point is important as it differs from the paradigm of generating the same distribution of the original dataset but rather a distribution that contains the information useful for downstream task training. To achieve this goal, a softmax probability matching method is proposed to minimize the KL-divergence between the softmax probabilities of real and synthetic data across a range of downstream model checkpoints, this range is needed to avoid the GAN memorizing the logits distribution of a single model. The proposed softmax matching loss (SML) enforces the generator $\boldsymbol{g}_\phi$ to generate representations that are useful for downstream model training. Specifically, let $\boldsymbol{\Theta}$ consist of a distribution of model checkpoints. For any sampled model $\boldsymbol{f}_\theta$ from this set, where $\theta \sim \boldsymbol{\Theta}$ represents the downstream model weights, the SML is defined as,
\begin{equation}
L_{\text{SML}} = \frac{1}{B}\sum_{j=1}^{B} \sum_{i=1}^{|\boldsymbol{y}|} \boldsymbol{f}_{\theta}(\mathbf{x_j})_i    \log\left(\frac{\boldsymbol{f}_{\theta}(\mathbf{x_j})_i}{\boldsymbol{f}_{\theta}(\boldsymbol{g}_\phi (\boldsymbol{z}_j) )_i}\right),
\label{eq:matching_loss}
\end{equation}

\noindent for $|\boldsymbol{y}|$, the number of classes, $B$ the batch size and $\boldsymbol{f}_{\theta}(\cdot)_i$ is the softmax probability of the i-th class given some real observation $\mathbf{x_j}$ or generated representation $\boldsymbol{g}_\phi(\boldsymbol{z}_j)$ . 

Furthermore, inspired by \cite{Yang2019DiversitySensitiveCG}, a diversity penalty is included into the generator $\boldsymbol{g}_{\phi}$ to encourage the generation of a wider variety of samples. Rather than producing samples clustered around a mode, the goal is to span the support of the real data as broadly as possible. The diversity penalty loss is defined as,
\begin{equation}
\resizebox{.90\linewidth}{!}{$
\text{L}_{\text{DIV}}(\boldsymbol{g}_\phi)= -\mkern-15mu \underset{\boldsymbol{z}1,\boldsymbol{z}2 \sim P(\boldsymbol{z})}{\mathbb{E}} \left[\min \left(\frac{ | \boldsymbol{g}_{\phi}\left(  \boldsymbol{z}1 \right) -\boldsymbol{g}_{\phi} \left( \boldsymbol{z}2 \right) | }{ | \boldsymbol{z}1 - \boldsymbol{z}2 |}, \tau\right)\right]$},
\label{eq:div_loss}
\end{equation}

\noindent with $\tau$ a scalar that bounds the diversity penalty for stability. Eq. \eqref{eq:div_loss} compares noises of the same class. Then, for two vectors $ \mathbf{z}_1$ and $\mathbf{z}_2$, if $\mathbf{z}_1  \approx \mathbf{z}_2$, the generator should generate two similar vectors. On the other hand, if the noises are different, then the generator should generate a different representation, thus avoiding mode collapse.

Finally, the proposed DD method consists of the same discriminator loss of Equation \eqref{eq:final_disc_before_dd} and the following generator loss,
\begin{equation}
L_{\text{G}_{\text{DD}}} = L_{\text{G}} + \lambda_5 \text{L}_{\text{DIV}} + \lambda_6 L_{\text{SML}}.
\label{eq:final_gen_dd}
\end{equation}

\section{Experiments}\label{sec:Experiments}
\subsection{Implementation details}\label{sec:diffteachers}
\vspace{-0.1cm}
\textbf{Dataset and SSL setup:} As explained in Section \ref{sec:introduction}, SER task is chosen and SUPERB \cite{superb} framework is followed for easy reproducibility and comparison with real data training. Experiments follow the leave-one-out session and only leave-out Session 1 is assessed due to computational resource restrictions. Nonetheless, in order to account for the possible variance in the results, McNemar's test is conducted at a p-value of 0.05 to verify statistical significance. The training data consists of Session 2 to 4, spanning 3,556 audios to distill. Motivated by \cite{wassertstein_gan_er_arousal_2018,gan_feature_dim_2018,cyclegan_transfer_emotion_2019} that does GAN data augmentation on a time averaged openSMILE \cite{opensmile} representation, this work generates a distribution on SSL representations but retaining the time dimension. Distillation is done over HuBERT Base \cite{hubert} SSL representations and evaluations are done with Unweighted Average Recall (UAR) to account for class imbalance.\\
\textbf{Discriminator and Generator Architecture:} Two small size architectures are considered. The first, named GAN-CNN, is a WGAN-GP model utilizing solely convolutional layers (CNN) for both its discriminator and generator. The discriminator is composed of 8 2D-CNN layers, each featuring layer normalization and leaky-relu activation. The final CNN layer connects to two feed-forward layers: one calculates the Wasserstein distance, and the other predicts the class category. The generator in GAN-CNN employs 2D CNN and transposed convolution layers followed by batch normalization and leaky-relu activation. There is no tanh operation at the generator's output, because the original SSL features are not limited to the [-1,1] range.

Using only CNN layers for the generator has the inductive bias that points that are spatially close to each other are correlated while neglecting long-range correlations. Nonetheless, for SER, modeling a long temporal context over each feature dimension may be important. Hence, the generator is modified to include only one self-attention operation over the time dimension after the 4-th CNN layer. To reduce number of parameters even further, the number of channels in the CNN layers are reduced from 256 to 128 and dilation is included to increase the receptive field of each CNN layer. This architecture is called GAN-ATT. Both models train on an A100-SXM4-40GB GPU, requiring 30 GPU hours each. Table \ref{tab:disk_storage_dd}, shows the sizes of the two GAN's architectures, highlighting a nearly 95\% size reduction compared to the original IEMOCAP audio. Such results in Table \ref{tab:disk_storage_dd} are important when scaling up to bigger datasets. 
\vspace{-0.1cm}

\begin{table}[t]
\setlength\tabcolsep{3pt}
\renewcommand{\arraystretch}{0.6}
\centering
\caption{DD size reduction of training set of IEMOCAP}
\begin{tabular}{l|c|c|c|c|c}
\toprule
           &  Audio  & SSL & GAN  & \multicolumn{2}{c}{Size Reduction  (\%)}    \\
           &   Files & feats &Size & Audio & SSL feats  \\
\midrule
GAN-CNN         &  \multirow{2}{*}{1.8 GB} &    \multirow{2}{*}{2.4 GB} &  0.1 GB  &  94.44 & 95.83\\%92.31 \\
GAN-ATT   &  &   &  0.06 GB &  96.95 & 97.50\\%95.38  \\
\bottomrule
\end{tabular}
\label{tab:disk_storage_dd}
\vspace{-0.1cm}
\end{table}
\vspace{-0.1cm}

\begin{table}[t]
\setlength\tabcolsep{0.47pt} % Adjust the horizontal padding here
\renewcommand{\arraystretch}{0.47} % Adjust the vertical padding here
\centering
\caption{SER UAR (\%$\uparrow$) for downstream model trained only with generated data. Two generators are evaluated, GAN-CNN and GAN-ATT, under 50 points per class (ppc) and 100 ppc. Baseline denotes the GAN without DD criterions. $\dagger$ denotes a McNemar's test statistically significant difference over the Baseline.$\ddagger$ denotes significance over the $+\text{L}_{\text{SML}}$ model.}
\label{tab:merged-results}
%\scalebox{0.88}{
\begin{tabular}{l|c|c|c|c|c|c}
\toprule
Model & \multicolumn{2}{c|}{GAN-CNN} & \multicolumn{4}{c}{GAN-ATT} \\
\cmidrule{1-7}
ppc & \multicolumn{1}{c|}{50} & \multicolumn{1}{c|}{100} & \multicolumn{1}{c|}{50} & \multicolumn{1}{c|}{100} & \multicolumn{1}{c|}{800} & \multicolumn{1}{c}{1800} \\
%\cmidrule{2-7}
%\cmidrule{2-9}
& UAR  & UAR  & UAR  & UAR & UAR & UAR  \\
\midrule
$\text{Baseline}$ & $47.75$ & $53.01$ & $56.31$ & $59.07$  & 61.44 & 62.05 \\
+ $\text{L}_{\text{DIV}}$ & $48.52$  & $53.25$  & $58.89$  & 60.52 & 62.20 & 62.86 \\
+ $\text{L}_{\text{SML}}$ & $53.79^{\dagger}$  & $60.52^{\dagger}$  & $56.99^{\dagger}$ & $60.26^{\dagger}$ & $63.96^{\dagger}$ & $64.07^{\dagger}$  \\
+ $\text{L}_{\text{DIV}}$ + $\text{L}_{\text{SML}}$ & $\textbf{54.99}^{\dagger \ddagger}$ & $\textbf{60.99}^{\dagger \ddagger}$ & $\textbf{60.27}^{\dagger \ddagger}$ & $\textbf{61.95}^{\dagger \ddagger}$ & $\textbf{64.35}^{\dagger \ddagger}$ & $\textbf{64.70}^{\dagger \ddagger}$\\
\bottomrule
\end{tabular}
%}
\end{table}

\vspace{-0.1cm}

\begin{table}[t!]
\setlength\tabcolsep{2pt}
\renewcommand{\arraystretch}{0.5}
\centering
\caption{UAR (\%$\uparrow$) test performance for real data training (Real SSL) and for GAN-ATT trained for balanced and imbalanced class labels distribution scenarios.}
\label{tab:best_result_condensed}
\begin{tabular}{l|c|c|c|c}
\toprule
Method &  2447 points & 2447 points & Full data & Full data \\
       &   Balanced & Imbalanced     &    Imbalanced & Balanced\\
\midrule
Real SSL  & 64.09           &  63.59   & 64.20 & - \\ %% Class accuracy 
GAN-ATT  &  64.47  &  63.21            & 63.69 & \textbf{64.50}  \\
\bottomrule
\end{tabular}
\end{table}

\subsection{GAN as a dataset distillator}
\vspace{-0.1cm}

Table \ref{tab:merged-results} analyzes the effect of training with a traditional WGAN-GP (Baseline) versus the proposed losses $\text{L}_{\text{DIV}}$ and $\text{L}_{\text{SML}}$ for the two generator architectures mentioned. DD aims to learn key discriminative information for training. In order to evaluate the efficiency of the discriminative information modelled, it is common to analyze DD performance using a small number of datapoints. Therefore, Table \ref{tab:merged-results} analyzes such results under a low points per class (ppc) budget of 50 ppc (5.6\% of the size of the original dataset) and 100 ppc (11.2\% of the size of the original dataset). Additionaly, to analyze how the proposal scales to bigger data samples, results with 800 ppc and 1800 ppc are included for GAN-ATT. All results are evaluated on the real data of Session 1 in IEMOCAP. From Table \ref{tab:merged-results}, it is evident that incorporating the proposed $\text{L}_{\text{DIV}}$ and $\text{L}_{\text{SML}}$ into the Baseline WGAN-GP significantly improves UAR across both generator architectures (GAN-CNN and GAN-ATT) and for both 50 ppc and 100 ppc dataset budgets. Furthermore, it can be seen that both terms are complementary. When comparing the Baseline with models using either $\text{L}_{\text{DIV}}$ or $\text{L}_{\text{SML}}$ individually, there is a noticeable improvement in performance. For instance, in the GAN-CNN architecture at 50 ppc, the UAR improves from 47.75\% to 48.52\% with $\text{L}_{\text{DIV}}$ and to 53.79\% with $\text{L}_{\text{SML}}$ and to 54.99\% when both terms are used together. For the GAN-ATT architecture, the trends are analogous. Interestingly, the GAN-ATT generator using both terms proposed for a budget of 11.2\% of the original dataset size can achieve an UAR score of 61.95\% which is only 2.25\% less than the model trained with the full original training data (see first row in Table \ref{tab:best_result_condensed}). For bigger data budgets, GAN-ATT surpasses the performance of the model trained with the original training set. Notably, two-tailed McNemar's test is performed at a p-value of 0.05 and results shows statistical significance for the $\text{L}_{\text{SML}}$ and $\text{L}_{\text{DIV}} + \text{L}_{\text{SML}}$ models when compared to the Baseline. Additionally, the $ \text{Baseline} + \text{L}_{\text{DIV}} + \text{L}_{\text{SML}}$ model is also statistically significant when compared with $+  \text{L}_{\text{SML}}$ only.

Table \ref{tab:best_result_condensed} compares GAN-ATT's performance against real data training in both balanced and imbalanced scenarios. IEMOCAP is a well known imbalanced dataset, meaning that some classes are represented more than others. Training with imbalanced data may hurt performance and hence using a GAN to alleviate this issue may be of importance. Last column of Table \ref{tab:best_result_condensed} shows that using the same size than the original dataset but with balanced classes improves performance than training with the original dataset. On the other hand, Full data Imbalanced column assess GAN-ATT under the same class label distribution of the original train set which shows similar performance than the model trained with real SSL. Besides, in order to test the real data set in a balance scenario, we select all the datapoints from the minority emotion class (693 utterances) and randomly select 693 utterances for each of the rest of emotion classes. Similarly, we analyze performance of 693 ppc for the GAN-ATT (2447 points in total) and finally we see performances of real SSL and GAN-ATT under the imbalance scenario for 2447 datapoints. Findings in Table \ref{tab:best_result_condensed} suggest that the proposed method can be used to alleviate data label imbalance because GAN only training can improve performance versus real data training. Such results suggest that having a generative model that can modify the train data class label distribution is beneficial and is a strength of this proposed method. Finally, we noticed that using GAN data makes the downstream model quickly converge on the real validation set, making the model to be trained in less than 5 minutes. On the other hand, real data training convergence is slower, taking around 90 minutes to train which is nearly a 95\% time reduction for downstream model training. This efficiency facilitates quicker hyperparameter optimization for downstream models, showcasing another advantage of our approach.
\vspace{-0.1cm}
\subsection{On the privacy aspect}
\vspace{-0.1cm}
Although this method does not inherently guarantee privacy, its use of GANs to learn SSL-like representations, conditioned solely on emotion labels, does not seem optimal to retain other forms of information.  This section focuses on speaker identity, but similar arguments can be made about the retention of information such as content. The model's design, results in the generation of abstract representations that enhance downstream model performance for SER. This implicitly bolsters privacy by limiting the frame-level information necessary for accurate speech reconstruction. To assess the potential for retaining speaker information, we propose testing using the downstream model's first layer as a speaker embedding, a technique widely recognized in speaker identification (SID) studies \cite{xvectors,liu2024disentangling}. Table \ref{tab:speaker_embeddings} shows such results for SUPERB SID task, where our proposed method reduces speaker information by 6.88\% compared to the linear layer of a downstream model trained for SER with real SSL representations. While these results do not mean the GAN-ATT ensures privacy, it does mean there is an implicit reduction on speaker identity modelling which could serve as a starting point for explicitly training DD that ensures privacy. This will be investigated in future work.

\begin{table}[t!]
\vspace{-0.1cm}
\setlength\tabcolsep{0.9pt}
\renewcommand{\arraystretch}{0.6}
\centering
\caption{SID Accuracy (Acc) with different speaker embeddings.}
\label{tab:speaker_embeddings}
\begin{tabular}{l|c}
\toprule
Speaker Embedding &  Acc (\%$\downarrow$) \\
\midrule
SID downstream model  & 80.89  \\ %% Class accuracy 
SER downstream model with Real SSL  & 44.87\\
SER downstream with Baseline WGAN-GP & 42.87 \\   
SER downstream with Proposed GAN-ATT	& \textbf{37.99} \\
\bottomrule
\end{tabular}
\vspace{-0.1cm}
\end{table}
\vspace{-0.1cm}

\section{Conclusions}

This study introduced DD for SER by leveraging a GAN to generate datasets that are useful for downstream model training. A softmax probability matching loss is proposed to achieve such goal. Diversity penalty is proposed to sample more variety of synthetic datapoints. The method achieves performance on par compared to real data downstream model training while substantially reducing dataset size and downstream training time. Our method can alleviate data label imbalance. Our method as well carries less speaker information which could serve as a starting point for an application on privacy preserving dataset distillation. Future work will analyze this direction as well as scaling to bigger datasets.

\section{Acknowledgments}
I want to deeply thank my friend Nikita Kuzmin for the great discussions on the privacy aspect. Unfortunately for him, I end up not adding such results on this manuscript.\\
The computational work for this article was fully performed on resources of the National Supercomputing Centre, Singapore (https://www.nscc.sg).

\bibliographystyle{IEEEtran}
\bibliography{mybib}

\end{document}